\documentclass{PoS}

\def\be{\begin{equation}}
\def\ee{\end{equation}}
\def\bea{\begin{eqnarray}}
\def\eea{\end{eqnarray}}
\def\ba{\begin{array}}
\def\ea{\end{array}}
\usepackage{diagbox}

\usepackage{amsmath}

\title{The many facets of exceptional field theory}

\ShortTitle{Exceptional field theory}

\author{Olaf Hohm$^a$ and Henning Samtleben$^b$\\
        \llap{$^a$} Institute for Physics, Humboldt University Berlin, Zum Gro{\ss}en Windkanal 6, D-12489 Berlin, Germany\\
         \llap{$^b$} Univ Lyon, Ens de Lyon, Univ Claude Bernard, CNRS,
Laboratoire de Physique, F-69342 Lyon, France \\
        E-mail: \email{ohohm@physik.hu-berlin.de}, \email{henning.samtleben@ens-lyon.fr}}

\abstract{Exceptional field theories are the manifestly duality covariant formulations of the target space theories of 
string/M-theory in the low-energy limit (supergravity) or for certain truncations.
These theories feature a rich system of sub-theories, corresponding to different consistent truncations. 
We review the structure of exceptional field theory and elaborate on various potential applications. 
In particular, we discuss how exceptional field theories capture some of the the magic supergravity triangles,
Hull's $*$-theories related to timelike dualities, as well as the generalized supergravities related to certain integrable deformations of AdS/CFT.
 }

\FullConference{Corfu Summer Institute 2018 "School and Workshops on Elementary Particle Physics and Gravity"\\
		(CORFU2018)\\
		31 August - 28 September, 2018\\
		Corfu, Greece}

\begin{document}

\section{Introduction}

There is mounting evidence for a unifying theory encompassing all 10-dimensional string theories 
and 11-dimensional supergravity, often referred to as `M-theory'. The defining features of this conjectured 
theory include  the so-called U-dualities, which are motivated by hidden symmetries that emerge upon 
toroidal compactification of maximal supergravity. Specifically, dimensionally reducing 11-dimensional supergravity 
on a $d$-dimensional torus leads to global symmetries in the exceptional series ${E}_{d(d)}(\mathbb{R})$, $d=2,\ldots, 8$,  (of non-compact split signature), 
leading in particular to the largest finite-dimensional exceptional group ${E}_{8(8)}(\mathbb{R})$ in three dimensions. 
M-theory is conjectured to be invariant under the discrete subgroups ${E}_{d(d)}(\mathbb{Z})$. The natural question arises 
of how to define a theory of (quantum) gravity that possesses such exceptional symmetries. 
Exceptional field theory, which is an extension of double field theory and defined on enlarged spaces with extended coordinates, 
provides, conservatively, a reformulation of supergravity  
that is manifestly U-duality covariant \textit{prior} to any compactification. More audaciously, 
exceptional field theory can be viewed as a first step towards a theory in between supergravity and the full string/M-theory that includes massive string or M-theory 
modes in duality-complete multiplets and that is a consistent truncation of the full M-theory.

Exceptional field theory (ExFT) as currently understood is defined in a Kaluza-Klein inspired `split formulation', 
with fields depending on `external' coordinates  $x^{\mu}$ and generalized `internal' coordinates $Y^M$ in the fundamental 
representation of the U-duality group ${E}_{d(d)}$ under consideration, for instance the adjoint representation for ${E}_{8(8)}$, where 
$M, N=1,\ldots, 248$. 
Moreover, the fields generically include the generalized metric ${\cal M}_{MN}$ parametrizing the coset space ${E}_{d(d)}/K(E_{d(d)})$, Kaluza-Klein-type 
vectors $A_{\mu}{}^{M}$, and an external metric $g_{\mu\nu}$. All fields depend on external and internal coordinates, 
with the dependence on the latter coordinates restricted by a duality-covariant `section constraint'. 
This constraint can be solved in various ways. Most simply, one may take the fields to be independent of the $Y^M$ 
coordinates, in which case one recovers directly the dimensionally reduced theory with enhanced global symmetries that originally motivated the 
notion of U-duality. 
More non-trivially, there are solutions of the section constraint that complete the external coordinates to the ten or eleven coordinates 
of type IIB or 11-dimensional supergravity. These theories are thereby reconstructed  in a split formulation that decomposes all coordinates and 
tensor components as in Kaluza-Klein but without truncating the coordinate dependence. In particular, the components of 
the generalized metric ${\cal M}_{MN}$ then encode the components of the internal Riemannian metric $g_{mn}$ and the 
internal components of various other tensor fields present in supergravity. 
In this way, ExFT provides a theory encompassing the complete, untruncated 10- or 11-dimensional supergravities, 
thereby rendering manifest the emergence of exceptional symmetries in lower dimensions.

While one can argue, under certain assumptions, that there are only two different `U-duality orbits' of solutions of the section 
constraint, leading to M-theory/type IIA or type IIB, respectively, we will use the opportunity  to illustrate 
how ExFT provides a rich set of consistent truncations and `alternative parametrizations' that give rise to novel applications. 
For instance, we will give ExFT in a formulation that is appropriate for timelike dualities. While in the conventional formulation the internal sector 
is taken to be Euclidean (in the sense that the internal metric $g_{mn}$ is Euclidean), one may also take the internal 
sector to be Lorentzian and the external sector to be Euclidean. The U-duality transformations, acting in particular on the generalized metric, 
then include timelike dualities. It should be emphasized that such dualities, while somewhat unconventional, are of physical interest since they apply 
whenever one has timelike Killing vectors (or BPS solutions). In addition, theories with a `Lorentzian generalized metric' are intriguing in that 
the latter may go through a phase in which the conventional metric components $g_{mn}$ become singular while the generalized metric 
is still perfectly regular. In this fashion, ExFT encodes truly `non-geometric' configurations.

Another application we will discuss here is the relation to the `magic tables' that arose in the literature since the early days 
of supergravity. Specifically, there is an abundance of interesting, often exceptional  groups arising in lower dimensions, 
with various intriguing interrelations and symmetries between them. Here we consider the `magic triangle' discovered by 
Cremmer et.~al.~in compactifications to three dimensions \cite{Cremmer:1999du}. We will show that the ${E}_{8(8)}$ ExFT unifies all theories 
corresponding to the different entries of that table into a single Lagrangian that, moreover, realises the symmetry of the triangle 
following a simple group-theory argument. 
Finally, we review how the generalized type IIB supergravity theory that appeared recently in integrable deformations of AdS/CFT 
 is naturally embedded in ExFT. 

The rest of this article is organized as follows. 
In sec.~2 we review ExFT with a focus on the ${E}_{6(6)}$ and ${E}_{8(8)}$ theories that will be employed later. 
In sec.~3 we introduce the magic triangle of Cremmer et.~al.~and explain how it is accommodated within  the ${E}_{8(8)}$ ExFT.
We then turn to timelike dualities and Hull's M$^*$-theories in sec.~4. In sec.~5 we discuss the embedding of generalized IIB supergravity 
theory, while we close with a brief outlook in sec.~6.

\section{Brief Review of Exceptional Field Theory}
\label{sec:review}

In this section we give a review of exceptional field theory (ExFT), 
originally constructed in \cite{Hohm:2013pua,Hohm:2013vpa,Hohm:2013uia,Hohm:2014fxa}.
We focus on the special theories corresponding to U-duality groups $E_{6(6)}$ (first subsection) 
and $E_{8(8)}$ (second subsection). For an extensive general review of ExFT we refer to the previous 
Corfu proceedings \cite{Baguet:2015xha}. For the lower-rank exceptional groups,
the associated ExFTs have been constructed in 
\cite{Hohm:2015xna,Abzalov:2015ega,Musaev:2015ces,Berman:2015rcc}.
The first example of an ExFT based on an infinite-dimensional duality group ($E_{9(9)}$)
is under construction~\cite{Bossard:2018utw}.

\subsection{E$_{6(6)}$}
We begin by reviewing some relevant  facts about the non-compact Lie group E$_{6(6)}$. 
It is 78-dimensional, with generators that we denote by  
$t_{\alpha}$ with the adjoint index $\alpha=1,\ldots, 78$. 
Apart from the adjoint representation, E$_{6(6)}$ carries two 
inequivalent fundamental representations of dimension $27$, 
denoted by ${\bf 27}$ and $\bar{\bf 27}$, which we label by lower indices 
$M,N=1,\ldots,27$ for ${\bf 27}$ 
and upper indices for $\bar{\bf 27}$. The (rescaled) Cartan-Killing form
can then be expressed in terms of generators $(t_{\alpha})_M{}^N$ in a 
fundamental representation as $\kappa_{\alpha\beta}\equiv (t_{\alpha})_M{}^N (t_{\beta})_N{}^M$. 
Moreover, the fundamental representations carry two completely symmetric invariant tensors 
of rank 3, the $d$-symbols $d^{MNK}$ and $d_{MNK}$, 
which are normalized as $d_{MPQ}d^{NPQ} = \delta_M^N$.
Finally, the tensor product ${\bf 27}\otimes \bar{\bf 27}$ contains the adjoint representation, 
and the corresponding projector onto the adjoint representation can be expressed explicitly as 
\bea\label{adjproj}
\mathbb{P}^M{}_N{}^K{}_L
&\equiv& (t_\alpha)_N{}^M (t^\alpha)_L{}^K ~=~
\frac1{18}\,\delta_N^M\delta^K_L + \frac16\,\delta_N^K\delta^M_L
-\frac53\,d_{NLR}d^{MKR}\;, 
\eea
satisfying $\mathbb{P}^M{}_N{}^N{}_M = 78$. 

We are now ready to develop the generalized geometry based on an extended spacetime with coordinates $Y^M$
in the ${\bf \bar{27}}$ of E$_{6(6)}$. All fields and gauge parameters appearing in the following will be functions of these 
coordinates, subject to a `section constraint' to be defined shortly. There is a notion of (infinitesimal) generalized 
diffeomorphisms, $Y^M\rightarrow Y^M-\Lambda^M(Y)$, acting on tensor fields 
according to generalized Lie derivatives. For instance, on a vector in the ${\bf \bar{27}}$  it reads 
\bea\label{genLie}
\delta V^M \ = \ \mathbb{L}_{\Lambda} V^M 
\ \equiv \ \Lambda^K \partial_K V^M - 6\, \mathbb{P}^M{}_N{}^K{}_L\,\partial_K \Lambda^L\,V^N
+\lambda\,\partial_P \Lambda^P\,V^M
\;, 
\eea
where $\lambda$ is arbitrary (intrinsic) density weight. 
The action of $\mathbb{L}_{\Lambda}$ on tensor fields in other representations of E$_{6(6)}$ follows similarly, 
such that all group-invariant operations are also invariant under generalized Lie derivatives. 
For instance, the $d$-symbols are gauge invariant, $\mathbb{L}_{\Lambda}d_{MNK} = 0$, 
and the action on the  E$_{6(6)}$-valued generalized metric ${\cal M}_{MN}$ preserves its group property. 
The generalized Lie derivatives form a closed algebra, 
 \be\label{Liealgebra}
  \big[\,\mathbb{L}_{\Lambda_1},\mathbb{L}_{\Lambda_2}\,\big] \ = \ \mathbb{L}_{[\Lambda_1,\Lambda_2]_{\rm E}}\;, 
 \ee 
governed by the so-called `E-bracket' 
\bea\label{Ebracket}
\big[\Lambda_1,\Lambda_2\big]^M_{\rm E} \ = \ 
2\Lambda_{[1}^K \partial_K \Lambda_{2]}^M
-10 \,d^{MNP}d_{KLP}\,\Lambda_{[1}^K \partial_N \Lambda_{2]}^L
\;, 
\eea
provided the functions defining the vector fields $\Lambda^M$ satisfy the section constraint 
 \bea
  d^{MNK}\partial_N\otimes \partial_K \ = \ 0\;. 
  \label{sectionE60}
 \eea
This somewhat symbolic presentation of the constraint is to be interpreted in the sense 
that for any functions $f, g$ on the extended spacetime 
 \bea\label{sectionE6}
  d^{MNK}\,\partial_N \partial_K f \ = \ 0\;, \quad  d^{MNK}\,\partial_Nf\, \partial_K g \ = \ 0 \,. 
 \eea  
These constraints are the M-theory version of the (stronger version of the) level-matching constraint in the string-inspired double field theory.

Having defined the gauge structure of generalized Lie derivatives, our next goal is to define a gauge 
theory based on it. Specifically, we introduce fields on an external five-dimensional spacetime with coordinates $x^{\mu}$, 
taking values in the internal generalized space with coordinates $Y^M$, subject to generalized diffeomorphisms. 
Thus, we consider fields and gauge parameters depending on coordinates $(x^{\mu}, Y^M)$. 
As a consequence, partial derivatives like $\partial_{\mu}$ will no longer be covariant under generalized Lie derivatives w.r.t.~parameters 
$\Lambda^M(x,Y)$, requiring the introduction of gauge covariant derivatives. 
We thus introduce gauge fields $A_{\mu}{}^{M}$ (of intrinsic density weight $\frac{1}{3})$ and covariant derivatives 
 \be\label{GeneralCOV}
  {\cal D}_{\mu} \ \equiv \ \partial_{\mu}-\mathbb{L}_{A_{\mu}}\;.
 \ee 
These derivatives transform covariantly under generalized diffeomorphisms, provided the 
gauge connections transform as  
 \bea\label{GaugeVar}
\delta A_\mu{}^M &\equiv & {\cal D}_\mu \Lambda^M
\;. 
\eea
Following the usual textbook treatment of gauge theories one would next define gauge covariant field strengths 
for $A_{\mu}{}^{M}$, but this turns out to be more subtle since the E-bracket does not actually define a Lie algebra. 
However, its failure to define a Lie algebra, and consequently the failure of the naive field strength to be gauge covariant, 
is of a controlled form, so that covariance can be repaired by introducing 2-forms $B_{\mu\nu M}$ in the ${\bf 27}$ 
(of intrinsic density weight $\frac{2}{3})$
and defining 
\bea
{\cal F}_{\mu\nu}{}^M &\equiv&
2\, \partial_{[\mu} A_{\nu]}{}^M  -\big[A_{\mu},A_{\nu}\big]^M_{\rm E} + 10 \, d^{MNK}\,\partial_K B_{\mu\nu\,N}
\;. 
\label{modF}
\eea
This curvature is fully gauge covariant, provided we assign appropriate gauge transformations to the 2-forms, 
a construction known as tensor hierarchy. Next, one can define a gauge covariant 3-form curvature for the 2-form field, 
which is implicitly determined by the hierarchical Bianchi identity 
 \be
  3 \,{\cal D}_{[\mu}{\cal F}_{\nu\rho]}{}^M \ = \ 10\, d^{MNK}\partial_K{\cal H}_{\mu\nu\rho\,N}\;. 
  \label{Bianchi}
 \ee 
(See \cite{Hohm:2019wql} for an extensive review of this `higher gauge theory' 
aspect of ExFT, \cite{{Hohm:2017pnh}} for a review of $L_{\infty}$ algebras, and \cite{Bonezzi:2019ygf} for a general theory of tensor hierarchies.)

We have now all structures in place in order to define the E$_{6(6)}$ ExFT \cite{Hohm:2013vpa}. Its field content is given by 
\bea
\left\{e_\mu{}^a, {\cal M}_{MN}, A_\mu{}^M, B_{\mu\nu\,M} \right\}
\;, 
\label{fieldcontent}
\eea
with $A_{\mu}$ and $B_{\mu\nu}$ the tensor fields introduced above, and where $e_\mu{}^a$ is the frame field 
of the five-dimensional external space (``f\"unfbein"), and ${\cal M}_{MN}$ is the generalized metric of the 27-dimensional 
internal generalized space. More precisely,  ${\cal M}_{MN}$ parameterizes the coset space ${\rm E}_{6(6)}/{\rm USp}(8)$. 
The f\"unfbein is a scalar density under internal generalized diffeomorphisms, 
with intrinsic density weight $\frac{1}{3}$. The generalized metric ${\cal M}_{MN}$ is an E$_{6(6)}$ valued symmetric tensor under 
generalized diffeomorphisms (of density weight zero). 
The complete bosonic action is given by 
\bea
 S_{\rm ExFT} &=&  \int dx\, dY\,e\, \Big( \widehat{R}
 +\frac{1}{24}\,g^{\mu\nu}{\cal D}_{\mu}{\cal M}^{MN}\,{\cal D}_{\nu}{\cal M}_{MN}
-\frac{1}{4}\,{\cal M}_{MN}{\cal F}^{\mu\nu M}{\cal F}_{\mu\nu}{}^N
 +e^{-1}{\cal L}_{\rm top}
-V(g,{\cal M})\Big) \,, 
\nonumber\\
\label{ExFTE6}
\eea
written in terms of the gauge covariant derivatives and field strengths defined above. Moreover, the various terms 
are defined as follows. The generalized Ricci scalar $\widehat{R}$ building  the Einstein-Hilbert term 
is defined by taking the familiar definition of the Ricci scalar and replacing every $\partial_{\mu}$ by ${\cal D}_{\mu}$, 
acting on $g_{\mu\nu}$ as a scalar density of weight $\frac{2}{3}$. The topological term can be defined concisely 
in terms of the curvatures (\ref{modF}), (\ref{Bianchi}) 
as a boundary integral in one higher dimension, 
\bea
S_{\rm top}\  = \
\kappa \!\int d^{27}Y \int_{{\cal M}_6}\,\left(
d_{MNK}\,{\cal F}^M \wedge  {\cal F}^N \wedge  {\cal F}^K
-40\, d^{MNK}{\cal H}_M\,  \wedge \partial_N{\cal H}_K
\right)
\,,
\label{CSlike}
\eea
where ${\cal M}_6$ is a six-manifold whose boundary is the (external) five-dimensional spacetime. 
The pre-factor is determined by (external) diffeomorphism invariance to be $\kappa= \sqrt{5/32}$. 
Finally, the last term in (\ref{ExFTE6}) --- sometimes referred to as `the potential', because it reduces upon 
compactification to the scalar potential of supergravity --- is defined by 
\be\label{fullpotential}
 \begin{split}
  V \ = \ &-\frac{1}{24}{\cal M}^{MN}\partial_M{\cal M}^{KL}\,\partial_N{\cal M}_{KL}+\frac{1}{2} {\cal M}^{MN}\partial_M{\cal M}^{KL}\partial_L{\cal M}_{NK}\\
  &-\frac{1}{2}g^{-1}\partial_Mg\,\partial_N{\cal M}^{MN}-\frac{1}{4}  {\cal M}^{MN}g^{-1}\partial_Mg\,g^{-1}\partial_Ng
  -\frac{1}{4}{\cal M}^{MN}\partial_Mg^{\mu\nu}\partial_N g_{\mu\nu}\;. 
 \end{split} 
 \ee 
Its relative coefficients are uniquely determined by (internal) generalized diffeomorphism invariance. 
This term can be written more geometrically, in terms of a generalized Ricci scalar ${\cal R}({\cal M}, g)$ of the internal geometry that, 
however, also depends on the external metric.

Being defined in terms of gauge covariant derivatives and field strengths, the above action is manifestly invariant under 
internal generalized diffeomorphisms with gauge parameters $\Lambda^M(x,Y)$. In particular, every term in (\ref{ExFTE6}) 
is separately invariant. On the contrary, 
all relative coefficients in (\ref{ExFTE6}) are fixed by external diffeomorphisms with parameters $\xi^{\mu}$. They act on the fields as 
 \bea
 \delta e_{\mu}{}^{a} &=& \xi^{\nu}{\cal D}_{\nu}e_{\mu}{}^{a}
 + {\cal D}_{\mu}\xi^{\nu} e_{\nu}{}^{a}\;, \nonumber\\
\delta {\cal M}_{MN} &=& \xi^\mu \,{\cal D}_\mu {\cal M}_{MN}\;,\nonumber\\
\delta A_{\mu}{}^M &=& \xi^\nu\,{\cal F}_{\nu\mu}{}^M + {\cal M}^{MN}\,g_{\mu\nu} \,\partial_N \xi^\nu
\;,\nonumber\\
\Delta B_{\mu\nu\,M} &=& \frac1{16\kappa}\,\xi^\rho\,
 e\varepsilon_{\mu\nu\rho\sigma\tau}\, {\cal F}^{\sigma\tau\,N} {\cal M}_{MN} 
 \;, 
 \label{skewD}
\eea
where we used the `covariant variations' of the 2-forms defined by 
\bea\label{DELTAB}
\Delta B_{\mu\nu\,M} &\equiv& \delta B_{\mu\nu\,M} + d_{MNK}\,A_{[\mu}{}^N\, \delta A_{\nu]}{}^K
\;.
\eea
Note that this invariance is manifest for gauge parameters $\xi^{\mu}$ depending only on $x^{\mu}$ coordinates, 
since all terms in the action are tensorial in the sense of the usual tensor calculus. 
However, invariance under parameters $\xi^{\mu}(x,Y)$ is no longer manifest, due to the presence of $\partial_M$
derivatives. 
The complete invariance is verified by a quite tedious computation, some aspects of which will be reviewed below, 
which employs in particular the following identity between the generalized metric and the $d$-tensor:
\bea
{\cal M}^{KL}{\cal M}^{MN}{\cal M}^{PQ} \,d_{LNQ}&=& d^{KMP} 
\;.
\label{MMMd}
\eea

In the remainder of this subsection we discuss how this theory relates to conventional supergravity, 
in particular how the section constraints are solved. In order to recover 11-dimensional supergravity 
in a $5+6$ split of coordinates we have to break E$_{6(6)}$ down to its subgroup
 \be
  {\rm SL}(6) \times {\rm GL}(1)_{{\rm 11}} \ \subset \  {\rm SL}(6) \times {\rm SL}(2) \ \subset \ E_{6(6)} \,,
 \ee 
with the fundamental representation of ${\rm E}_{6(6)}$ breaking as
\bea
{\bf 27} &\longrightarrow& 6_{+1} + 15'_{0} + 6_{-1} \;, 
\eea
with the subscripts referring to the ${\rm GL}(1)$ charges.
For the coordinates $Y^M$ in this representation this corresponds to the decomposition 
\bea
\left\{Y^M\right\} &\longrightarrow&\left\{\, y^m\,,\; y_{mn}\,, \; y^{\bar{m}} \,\right\}
\;. 
\label{break11D}
\eea
An explicit solution to the section condition (\ref{sectionE6}) is given by 
restricting the $Y^M$ dependence of all fields to the six coordinates $y^m$\,, 
as can be seen by working out the $d$-tensor in this decomposition, see  \cite{Hohm:2013vpa} for details.

The type IIB supergravity theory in a $5+5$ split of coordinates is recovered upon breaking E$_{6(6)}$ down to its subgroup
 \bea
   {\rm SL}(5)\times{\rm SL}(2)\times{\rm GL}(1)_{\rm IIB}\ \subset \ E_{6(6)}\;, 
 \eea  
such that
\bea
{\bf 27} &\longrightarrow&
(5,1)_{+4} + (5',2)_{+1} +(10,1)_{-2} + (1,2)_{-5}
\;,
\eea
and thus for the coordinates 
\bea
\left\{Y^M\right\} &\longrightarrow& \{y^a\,,\; \tilde{y}_{a\alpha}\,,\;  \tilde{y}^{ab}\,,\;  \tilde{y}_{\alpha} \}
\;,\qquad
a=1, \dots, 5\;,\quad \alpha=\pm
\;. 
\label{breakIIB}
\eea
The section constraint is then solved by restricting the internal coordinate dependence of all fields to the coordinates $y^a$\,.
This solution is inequivalent to (\ref{break11D}) in the sense that no E$_{6(6)}$ rotation can map the $y^a$
coordinates from (\ref{breakIIB}) into (a subset of) the $y^m$ coordinates from (\ref{break11D}).

We close with a few remarks on the dictionary between the ExFT fields and the conventional supergravity fields. 
To this end one must pick a solution of the section constraint and a group decomposition as above. 
For instance, under the decomposition (\ref{break11D}) appropriate for $D=11$ supergravity 
the generalized metric given by the symmetric matrix ${\cal M}_{MN}$ decomposes 
into the block form
\bea
{\cal M}_{KM} &=& \left(
\begin{array}{ccc}
{\cal M}_{km}&{\cal M}_k{}^{mn}&{\cal M}_{k\bar{m}}\\
{\cal M}^{kl}{}_m & {\cal M}^{kl,mn} & {\cal M}^{kl}{}_{\bar{m}}\\
{\cal M}_{\bar{k}m}& {\cal M}_{\bar{k}}{}^{mn}&{\cal M}_{\bar{k}\bar{m}}
\end{array}
\right)
\;.
\label{M27}
\eea
These components should now be expressed in terms of supergravity fields. 
For instance, its last line reads
\bea
{\cal M}_{\bar{m}n} &=& \frac1{24} \,g^{-2/3}  g_{mk} \,\epsilon^{klpqrs} \,c_{nlp}c_{qrs}-g^{-2/3}  g_{mn}\,\varphi
\;,
\nonumber\\
{\cal M}_{\bar{m}}{}^{kl} &=& -\frac1{6\,\sqrt{2}}\,g^{-2/3}\,g_{mn} \epsilon^{nklpqr}\, c_{pqr}\;,
\qquad
{\cal M}_{\bar{m}\bar{n}} ~=~ g^{-2/3}\, g_{mn}\;,
\label{compM1}
\eea
parametrized by $\varphi$, $c_{mnk}$ and the six-dimensional internal Euclidean metric $g_{mn}$
with determinant $g\equiv{\rm det}\,g_{mn}$\,. 
The internal metric originates directly from the Kaluza-Klein-type decomposition of the 11-dimensional metric. 
The antisymmetric tensor $c_{mnk}$ encodes the internal components of the 11-dimensional 3-form field, 
while the scalar $\varphi$ is the dualization (in five dimensions) of the purely external components of the 11-dimensional 3-form field. 
The dictionary (\ref{compM1}) is straightforwardly established by comparing the action of generalized diffeomorphisms (\ref{genLie})
onto the various blocks of (\ref{M27}) to the internal diffeomorphism and $p$-form gauge transformations of the higher-dimensional
supergravity fields.
The remaining blocks of (\ref{M27})
can be expressed in compact form via the matrix
\bea
\tilde{\cal M}_{MN} &\equiv& {\cal M}_{MN}- {\cal M}_{M\bar{m}} ({\cal M}_{\bar{m}\bar{n}})^{-1} {\cal M}_{\bar{n}N}
\;,
\label{tildeM27}
\eea
which take the form
\bea
\tilde{\cal M}_{mn}  &=& g^{1/3}\,g_{mn} + \frac12\,g^{1/3}\, c_{mkp}c_{nlq} \,g^{kl} g^{pq}
\;,\nonumber\\
\tilde{\cal M}_{m}{}^{kl}  &=& -\frac1{\sqrt{2}}\,g^{1/3}\,c_{mpq} \,g^{pk} g^{ql}
\;,\qquad
\tilde{\cal M}^{kl,mn}  ~=~ g^{1/3}\,g^{m[k}g^{l]n}
\;.
\label{compM2}
\eea
The type IIB dictionary takes an analogous form
based on the block decomposition of the symmetric matrix ${\cal M}_{MN}$
based on the decomposition (\ref{breakIIB}).
Details are spelled out in \cite{Baguet:2015xha}.

\subsection{E$_{8(8)}$}

The exceptional field theory based on the Lie group E$_{8(8)}$ has been constructed in~\cite{Hohm:2014fxa}.
It is built starting from the reduction of eleven-dimensional supergravity down to three dimensions
which exhibits a global E$_{8(8)}$ symmetry.
Like all exceptional field theories with only three external dimensions, the construction of this
theory brings about some particular features, related to the Ehlers type symmetry enhancement \cite{Ehlers:1957}
which  necessitates to describe within the scalar sector some of the dual graviton degrees of freedom.

Vector fields in the E$_{8(8)}$ ExFT transform in the adjoint ${\bf 248}$ representation of E$_{8(8)}$.
Accordingly, generalized diffeomorphisms take the form analogous to (\ref{genLie})
\bea\label{genLieE8A}
\delta_\Lambda V^M \ = \ \mathbb{L}_{\Lambda} V^M 
\ \equiv \ \Lambda^K \partial_K V^M - 60\, \mathbb{P}^M{}_N{}^K{}_L\,\partial_K \Lambda^L\,V^N
+\lambda\,\partial_P \Lambda^P\,V^M
\;, 
\eea
with indices $M, N, \dots$ now labelling the adjoint representation of E$_{8(8)}$,
raised and lowered by the Cartan-Killing form $\eta_{MN}$, and the
projector $\mathbb{P}^M{}_N{}^K{}_L =
\frac1{60}\,f^M{}_{NP}\,f^{PK}{}_L{}$ onto the adjoint representation.
In contrast to the lower rank exceptional field theories, the transformations (\ref{genLieE8A})
no longer close into an algebra among themselves but give rise to an enhanced symmetry algebra
which also carries covariantly constrained local E$_{8(8)}$ rotations as
\bea\label{genLieE8B}
\delta_\Sigma V^M \ = \ -\Sigma_K \,f^{KM}{}_N\, V^N 
\;,
\eea
with the structure constants $f^{MN}{}_K$.
The full algebra (\ref{genLieE8A}), (\ref{genLieE8B}) now closes according to
\bea\label{closure}
  \big[\,\delta_{(\Lambda_{1},\Sigma_1)},\,\delta_{(\Lambda_{2},\Sigma_2)}\,\big] &=& 
  \delta_{[(\Lambda_{2},\Sigma_2),(\Lambda_{1},\Sigma_1)]_{\rm E}}\;, 
 \qquad
 [(\Lambda_{2},\Sigma_2),(\Lambda_{1},\Sigma_1)]_{\rm E} ~\equiv~ (\Lambda_{12},\Sigma_{12})
 \;,\quad
 \eea
with the effective parameters
 \bea\label{EFFimp}
  \Lambda^M_{12} &\equiv&
2\,\Lambda_{[2}^N\partial_N \Lambda_{1]}^M
-14\,  (\mathbb{P}_{3875}){}^{MK}{}_{NL} \,\Lambda_{[2}^N \partial_K \Lambda_{1]}^L
-\frac1{4}\,\eta^{MK}\eta_{NL}\,\Lambda_{[2}^N \partial_K \Lambda_{1]}^L\nonumber\\
&&{}
+\frac1{4}\,f^{MN}{}_{P}\,\partial_N(f^{P}{}_{KL}\Lambda_2^K  \Lambda_1^L)
\;,\nonumber\\[1ex]
\Sigma_{12\,M} &\equiv&
-2\,\Sigma_{[2\,M} \partial_N \Lambda_{1]}^N 
+2\,\Lambda_{[2}^N \partial_N \Sigma_{1]\,M} 
-2\, \Sigma_{[2}^N \partial_M \Lambda_{1]\,N}
+f^N{}_{KL} \,\Lambda_{[2}^K\,\partial_M\partial_{N}\Lambda_{1]}^L
\;,\qquad
 \eea 
with $(\mathbb{P}_{3875}){}^{MK}{}_{NL}$ denoting the projector of the symmetric
product ${\bf 248}\otimes{\bf 248}$ onto the ${\bf 3875}$ representation.
The closure (\ref{closure})  requires the E$_{8(8)}$ section contraints
 \be\label{secconstr}
  \eta^{MN}\partial_M\otimes  \partial_N \ = \ 0 \;, \quad 
  f^{MNK}\partial_N\otimes \partial_K \ = \ 0\;, \quad 
  \big(\mathbb{P}_{3875}\big)_{MN}{}^{KL} \partial_K\otimes \partial_L \ = \ 0
  \;,
 \ee 
in analogy with (\ref{sectionE60}) for the internal coordinate dependence, accompanied by
the algebraic constraints
 \be
  \big(\mathbb{P}_{1+248+3875}\big)_{MN}{}^{KL} \,\Sigma_K\otimes \partial_L \ = \ 0 \ = \
    \big(\mathbb{P}_{1+248+3875}\big)_{MN}{}^{KL} \, \Sigma_K\otimes \Sigma_L
  \;,
  \label{section_full}
 \ee 
for the symmetry parameter of (\ref{genLieE8B}).

The field content of the E$_{8(8)}$ ExFT comprises scalar fields, parametrizing the coset space
${\rm E}_{8(8)}/{\rm SO}(16)$ in terms of the symmetric positive matrix ${\cal M}_{MN}$,
together with gauge fields $A_\mu{}^M$, $B_{\mu\,M}$, associated to the internal
diffeomorphisms (\ref{genLieE8A}), (\ref{genLieE8B}) with minimal couplings via covariant derivatives
 \be\label{covder}
  {\cal D}_{\mu} \ \equiv \ \partial_{\mu}-{\mathbb{L}}_{(A_{\mu},B_{\mu})}\;, 
 \ee 
 to the scalar fields.
The full action reads
\bea
S_{\rm ExFT} &=&  \int dx\, dY\,e\, \Big( \widehat{R}+\frac{1}{48}\,g^{\mu\nu}{\cal D}_{\mu}{\cal M}^{MN}\,{\cal D}_{\nu}{\cal M}_{MN}
  + \frac12\,e^{-1}\,{\cal L}_{\rm CS} - V(g, {\cal M})\Big)\;.
  \label{E8action}
\eea
with Einstein-Hilbert term and scalar kinetic term defined as in (\ref{ExFTE6}) above. The 
non-abelian Chern-Simons term is most conveniently defined as the boundary contribution of a 
manifestly gauge invariant exact form in four dimensions.  
\bea 
   S_{\rm CS} &\propto& \int_{\Sigma_4} d^4 x \,\int d^{248} Y\,
   \Big({\cal F}^M\wedge {\cal G}_M-\frac{1}{2}f_{MN}{}^{K} {\cal F}^M\wedge \partial_K {\cal F}^N \Big)
 \;,
    \label{CS4}
\eea
in terms of the non-abelian field strengths ${\cal F}_{\mu\nu}{}^M$, ${\cal G}_{\mu\nu\,M}$
associated to the gauge fields $A_\mu{}^M$, $B_{\mu\,M}$. We refer to~\cite{Hohm:2014fxa}
for their explicit definitions and to \cite{Hohm:2019wql} for an interpretation in terms of 
Leibniz gauge theories.
The potential term $V({\cal M},g)$ is finally given by
\bea\label{PotIntro}
  V({\cal M},g) & = &
  -\frac{1}{240}{\cal M}^{MN}\partial_M{\cal M}^{KL}\,\partial_N{\cal M}_{KL}+
  \frac{1}{2}{\cal M}^{MN}\partial_M{\cal M}^{KL}\partial_L{\cal M}_{NK} \\
  &&{}
  +\frac1{7200}\,f^{NQ}{}_P f^{MS}{}_R\,
  {\cal M}^{PK} \partial_M {\cal M}_{QK} {\cal M}^{RL} \partial_N {\cal M}_{SL} 
  \nonumber\\
  &&{}
  -\frac{1}{2}g^{-1}\partial_Mg\,\partial_N{\cal M}^{MN}-\frac{1}{4}  {\cal M}^{MN}g^{-1}\partial_Mg\,g^{-1}\partial_Ng
  -\frac{1}{4}{\cal M}^{MN}\partial_Mg^{\mu\nu}\partial_N g_{\mu\nu}\;,
  \nonumber
\eea
generalizing the potentials of the lower-rank exceptional fields theories such as (\ref{fullpotential})
by a new contribution explicitly carrying the ${\rm E}_{8(8)}$ structure constants.
Again, every term in the action (\ref{E8action}) is separately invariant under generalized internal
diffeomorphisms, while the relative coefficients are uniquely fixed by imposing invariance under
external diffeomorphisms. Upon solving the section constraints and implementing the proper dictionary,
the field equations of (\ref{E8action}) again reproduce those of the full $D=11$ and IIB supergravity, respectively.

\section{The magic triangle}

In this section we consider an instance of the `magic tables' that appeared in the supergravity literature since the discovery of maximal supergravity 
in eleven dimensions. As briefly recalled in the introduction, 11-dimensional supergravity has the intriguing property of giving rise to the 
exceptional symmetry groups ${E}_{d(d)}$ in compactifications,  on a $d$-torus, to $D=11-d$ dimensions. 
In particular, in three dimensions ($D=3$) one encounters the largest of the finite-dimensional exceptional groups, ${E}_{8(8)}$, 
which is the global symmetry of (ungauged) maximal three-dimensional supergravity whose propagating bosonic degrees of freedom are entirely 
encoded in the scalar fields of a non-linear sigma model based on ${E}_{8(8)}/{\rm SO}(16)$. 

Irrespective of their origin in maximal supergravity one may consider such $G/H$ coset models coupled to (topological) three-dimensional 
gravity in their own right. The action reads 
 \be
  S\ = \ \int d^3x\,\sqrt{g}\Big(R \ - \ P^{\mu A} P_{\mu}{}^{A}\Big)\;, 
 \ee
with the Einstein-Hilbert term for the three-dimensional metric $g_{\mu\nu}$ and a scalar kinetic term 
for the coset currents $P_{\mu}{}^{A}\equiv [{\cal V}^{-1}\partial_{\mu}{\cal V}]^A$, where ${\cal V}$ is a $G$-valued 
field and $[\cdot]$ denotes the projection onto the non-compact part (the complement of $H$), labelled 
by indices $A=1,\ldots, {\rm dim}(G)-{\rm dim}(H)$. 
We may consider the coset models of the exceptional series in $D=3$, as done by Cremmer 
et.~al.~\cite{Cremmer:1999du}, who also gave a systematic analysis of the higher dimensional origin of these models. 
(See also the earlier work in \cite{Breitenlohner:1987dg} and the subsequent completions in~\cite{Keurentjes:2002xc,Keurentjes:2002rc}.) 
The various higher-dimensional theories that these three-dimensional models can be uplifted to are indicated  in the table below. 
The columns are labelled by the rank of the group in $D=3$, 
and the rows are labelled by the highest dimension to which the three-dimensional models can be uplifted to. 
%\bea

{\footnotesize
\begin{equation}\nonumber 
\begin{tabular}{c||c|c|c|c|c|c|c|c|c|c|c|c|c|c|}
$11$ & {\scriptsize $\times$ }
\\
$10$ &
{\scriptsize $\;\mathbb{R}$} &  {\scriptsize ${\rm A}_{1}$}& {\scriptsize $\times$}
\\
$9$ &
\multicolumn{2}{c|}{${\rm A}_{1}\times \mathbb{R}$}
& {\scriptsize  $\mathbb{R}$}
\\
$8$ &
\multicolumn{2}{c|}{${\rm A}_{2}\times {\rm A}_{1}$}
& 
{\scriptsize $ {\rm A}_{1}\times \mathbb{R}$} &  {\scriptsize ${\rm A}_{2}$}
& ${\rm A}_{1}$
\\
$7$ &
\multicolumn{2}{c|}{${\rm A}_{4}$}
&
\multicolumn{2}{c|}{${\rm A}_{2}\times \mathbb{R}$}
&$ {\rm A}_{1}\times \mathbb{R}$
& 
 {\scriptsize $\mathbb{R}$} &
& {$\times$}
\\
$6$ &
\multicolumn{2}{c|}{${\rm D}_{5}$}
&
\multicolumn{2}{c|}{${\rm A}_{3}\times {\rm A}_{1}$}
&${\rm A}_{1}^2\times \mathbb{R}$
&{\scriptsize $\mathbb{R}^2$ }& {\scriptsize ${\rm A}_{1}^2$}
&
$\mathbb{R}$
\\
$5$ &
\multicolumn{2}{c|}{${\rm E}_{6}$}
&
\multicolumn{2}{c|}{${\rm A}_{5}$}
&
${\rm A}_{2}^2$ &
\multicolumn{2}{c|}{$ {\rm A}_{1}^2\times \mathbb{R}$}
&$ {\rm A}_{1}\times \mathbb{R}$
& {\scriptsize ${\rm A}_{1}$}
\\
$4$ &
\multicolumn{2}{c|}{${\rm E}_{7}$}
&
\multicolumn{2}{c|}{${\rm D}_{6}$}
&
${\rm A}_{5}$&
\multicolumn{2}{c|}{${\rm A}_{3}\times {\rm A}_{1}$}
&
$ {\rm A}_{2}\times \mathbb{R}$&
{\scriptsize ${\rm A}_{1}\times \mathbb{R}$} &  {\scriptsize ${\rm A}_{2}$}&
$\mathbb{R}$ &
{\scriptsize $\times$}&\\
$3$ &
\multicolumn{2}{c|}{${\rm E}_{8}$}
&
\multicolumn{2}{c|}{${\rm E}_{7}$}
&
${\rm E}_{6}$&
\multicolumn{2}{c|}{${\rm D}_{5}$}
&
${\rm A}_{4}$&
\multicolumn{2}{c|}{${\rm A}_{2}\times {\rm A}_{1}$}
&
${\rm A}_{1}\times \mathbb{R}$&
{\scriptsize ${\rm A}_{1}$}&
{\scriptsize $ \mathbb{R}$} & $\times$\\
\hline
%\slashbox{$D$}{$r$} 
\diagbox[dir=SW,height=2.2em]{$D$}{$r$}
&
\multicolumn{2}{c|}{8}&
\multicolumn{2}{c|}{7}&6&
\multicolumn{2}{c|}{5}&4&
\multicolumn{2}{c|}{3}&2&
\multicolumn{2}{c|}{1}&0
\end{tabular}
\end{equation}
%\eea
}

Let us now discuss some features of this triangle. The first row ($D=3$) displays the `initial data', 
the exceptional series of global symmetries under consideration. The first column ($r=8$) displays the 
symmetry groups in all dimensions from $D=3$ to $D=11$. By construction, the groups of the first row 
equal the groups of the first column, for the groups in the first row were chosen in this way.  
However, remarkably this symmetry extends to the entire triangle (justifying its `magic') 
in that under the exchange of rank and dimension according to 
 \be\label{rDduality}
  (r,D) \ \leftrightarrow \ (11-D, 11-r)
 \ee
the groups stay the same. This is intriguing, for a priori the corresponding theories have little to do 
with each other, typically  being defined in completely different dimensions.  
This raises the question whether there is an overarching framework that not only 
explains this `duality' but also provides a theory from which the models corresponding to the different entries 
of the triangle can be derived by suitable truncations.

In the following we will explain that precisely the E$_{8(8)}$ ExFT provides this framework. 
In this we will use a group-theoretical argument anticipated by Keurentjes some time ago \cite{Keurentjes:2002xc} that
unfolds its full force in the context of ExFT. 
This group-theoretical argument relies on the fact that E$_{8(8)}$ can be decomposed according to 
\bea\label{E8Breaking}
{\rm E}_{8(8)}
&\longrightarrow&
{\rm SL}(D-2) \times {\rm SL}(9-r) \times U_{D,r}
\;,
\eea
where $U_{D,r}$ is the U-duality group labelled by $(D,r)$ in the table. 
In this sense, $U_{D,r}$ can be obtained from E$_{8(8)}$ by singling out two SL$(n)$ factors. 
There is, of course, no intrinsic difference between these two factors, and since they are 
interchanged under the duality (\ref{rDduality}) we infer the symmetry $U_{D,r}=U_{11-r,11-D}$~\cite{Keurentjes:2002xc}, 
thereby explaining this equality of groups, see also \cite{HenryLabordere:2002dk}.

More intriguingly, the E$_{8(8)}$ ExFT provides a unifying theory from which all entries of the table can be obtained 
through truncation, following the breaking of E$_{8(8)}$ displayed in (\ref{E8Breaking}). 
Specifically, one then decomposes the internal coordinates in the adjoint representation 
according to 
\bea
\left\{
Y^M
\right\} &\rightarrow&
\left\{
Y^{\underline i}{}_{\underline j}\,, \;Y^a{}_b, \dots
\right\}
\;,
\eea
where ${\underline i}, {\underline j}=1,\ldots, D-2$ label SL$(D-2)$ indices 
and $a, b = 1,\ldots, 9-r$ label ${\rm SL}(9-r)$ indices.  We displayed the coordinates in 
the adjoint of SL$(D-2)$ and ${\rm SL}(9-r)$, while the ellipsis indicates the remaining coordinates 
transforming non-trivially under $U_{D,r}$. 
In the next step one decomposes ${\underline i}=(i,\underline{0})$, where $i=1,\ldots, D-3$ is an SL$(D-3)$ index, 
and identifies  the $D-3$ internal physical coordinates as 
 \be
   y^i \ \equiv \  Y^{i}{}_{\underline{0}}\;, 
 \ee
which solves the E$_{8(8)}$ section constraint.
Together with the three external coordinates $x^{\mu}$, these describe the $D$-dimensional 
space-time on which the theory with duality group $U_{D,r}$ is defined.
In order to obtain this theory one, finally, truncates the E$_{8(8)}$ ExFT to singlets under ${\rm SL}(9-r)$. 
Since it is always consistent to truncate to the singlets of a symmetry group, this yields a consistent truncation. 
It would be interesting to see if these consistent truncations can be extended to gauged supergravities, 
using the techniques of generalized Scherk-Schwarz compactifications \cite{Lee:2014mla,Hohm:2014qga,Baguet:2015sma}
and to see to which extent the magic of the triangle is inherited by its gauged deformations.
It would also be interesting to explore to which extent other magic triangles \cite{Julia:1982gx} and pyramides
\cite{Anastasiou:2017taf} can be accommodated in ExFT. These typically feature non-split forms of the
exceptional groups. Intriguingly, the ExFT construction is largely independent of the particular real form used.
The latter is only visible via the solution space of the section constraints and via the parametrization of the scalar matrix ${\cal M}_{MN}$.

\section{Timelike dualities}

As reviewed above, the E$_{d(d)}$ ExFT Lagrangian is modelled after the theory
obtained by toroidal reduction of 11-dimensional supergravity on a spacelike torus $T^d$
in such a way that it still includes the full 11-dimensional supergravity.
In particular, is is based on the scalar coset space E$_{d(d)}/{\rm K}_d$ with K$_d$ 
the compact subgroup of E$_{d(d)}$.
It is then natural to expect an equivalent ExFT formulation of the higher-dimensional
theories based on their toroidal reduction on a torus including a timelike circle.
These reductions to Euclidean theories have been studied in~\cite{Stelle:1998xg,Hull:1998br,Cremmer:1998em}. 
They are closely related to spatial reductions, however, based on coset spaces E$_{d(d)}/\tilde{\rm K}_d$ 
with $\tilde{{\rm K}}_d$ now representing a different real form of K$_d$. This accounts for the fact
that timelike dimensional reduction in particular results in certain sign flips in the 
lower-dimensional kinetic terms. 
Extending these theories to their full ExFT form thus gives rise to yet an alternative
formulation of 11-dimensional supergravity based on a Euclidean external space and
a Lorentzian exceptional geometry. 

Moreover, by construction these theories should (and do) also accommodate Hull's $*$-theories \cite{Hull:1998vg,Hull:1998ym},
defined by T-duality along timelike isometries, as well as the underlying 11-dimensional supergravities of exotic signature. 
More precisely, T-duality along a timelike circle 
maps IIA and IIB supergravity into the so-called IIB$^*$ and IIA$^*$ supergravity, respectively, which differ from their 
unstarred counterparts by certain sign flips in the kinetic terms. Under timelike dimensional reduction, all these
theories give rise to the same lower-dimensional Euclidean supergravities based on the coset spaces from the
E$_{d(d)}/\tilde{\rm K}_d$ series.
Then it does not come as a surprise that the duality covariant ExFT formulations based on these coset spaces do
indeed accommodate in a single framework IIA/IIB supergravity together with their IIA$^*$/IIB$^*$ counterparts.

In this section, we illustrate these structures for the Euclidean version of E$_{6(6)}$ ExFT.
Hull's $*$-theories appeared earlier in type II double field theory \cite{Hohm:2011zr,Hohm:2011dv}.
For the internal sector of ${\rm SL}(5)$ ExFT these scenarios have also been discussed in~\cite{Blair:2013gqa},
then completed in \cite{Berman:2019izh}, see also \cite{Malek:2013sp} for earlier work.

\subsection{E$_{6(6)}$ ExFT with Euclidean external space}

Reduction of 11-dimensional supergravity on a 6-torus $T^{5,1}$ including a timelike circle leads
to a Euclidean five-dimensional theory based on the coset space ${\rm E}_{6(6)}/{\rm USp}(4,4)$
of indefinite signature~\cite{Stelle:1998xg,Hull:1998br,Cremmer:1998em}. 
More precisely, under ${\rm USp}(4)\times {\rm USp}(4)$ the 42 physical scalars parametrizing this
coset space decompose according to
\bea
{\bf 42} &\longrightarrow& (5,5)+(1,1)+(4,4)
\;,
\eea
with the $(4,4)$ corresponding to 16 compact directions within ${\rm E}_{6(6)}$, 
thus carrying an opposite sign in the kinetic sigma-model term. 
A quick counting shows that $16=5+10+1$ is indeed the number of scalars which flip the
sign of their kinetic term under reduction including a timelike circle: reduction of pure
gravity on $T^{5,1}$ yields a coset space ${\rm GL}(6)/{\rm SO}(5,1)$ with 
5 compact directions, the 11D three-form adds 10 scalars $C_{0mn}$ of opposite kinetic
term, as well as a three-form $C_{\mu\nu\rho}$, which in 5 dimensions is dualized to a scalar
of opposite kinetic sign (since the metric $g_{\mu\nu}$ is now Euclidean).

It is straightforward to adapt the ExFT construction to this setting. First, we note that
the structure of generalized diffeomorphisms including the full tensor hierarchy depends only on the 
algebra of ${\rm E}_{6(6)}$ and therefore remains identical to the previous construction.
The alternative ${\rm E}_{6(6)}$ ExFT Lagrangian can then be constructed from first principles and
 takes a form which is formally identical to ({\ref{ExFTE6})
\bea
 \tilde{S}_{\rm EFT} &=&  \int dx\, dY\,e\, \Big( \widehat{R}
 +\frac{1}{24}\,g^{\mu\nu}{\cal D}_{\mu}{\cal M}^{MN}\,{\cal D}_{\nu}{\cal M}_{MN}
-\frac{1}{4}\,{\cal M}_{MN}{\cal F}^{\mu\nu M}{\cal F}_{\mu\nu}{}^N
 +e^{-1}{\cal L}_{\rm top}
-V({\cal M},g)\Big) \,,
\nonumber\\
\label{ExFTEuc}
\eea
and only differs in the signatures of external and internal metric. Specifically,
in contrast to ({\ref{ExFTE6}), the external five dimensions now come with a Euclidean metric $g_{\mu\nu}$
while the internal metric ${\cal M}_{MN}$ parametrizing the coset space ${\rm E}_{6(6)}/{\rm USp}(4,4)$,
is no longer positive definite, but of signature $(16,11)$.
In particular, it satisfies the relation
\bea
{\cal M}^{KL}{\cal M}^{MN}{\cal M}^{PQ} \,d_{LNQ}&=& -d^{KMP} 
\;,
\label{MMMdEuc}
\eea
which differs in sign from the relation (\ref{MMMd}) satisfied on the non-compact coset space.

As in the construction of ({\ref{ExFTE6}), each term in (\ref{ExFTEuc}) is uniquely fixed by invariance
under generalized diffeomorphisms (which are blind to issues of space-time signature), while
the relative factors between the five terms are fixed by invariance of the action under external diffeomorphisms. 
The latter are given by the same transformation rules (\ref{skewD}) with a single change 
of sign in the transformation of two-forms, 
\bea
\Delta B_{\mu\nu\,M} &=& -\frac{1}{16\kappa}\,\xi^\rho\,e\,\varepsilon_{\mu\nu\rho\sigma\tau}\,{\cal F}^{\sigma\tau\,N}\,{\cal M}_{MN}
\;,
\label{DBEuc}
\eea
required such as to achieve the standard on-shell transformation
\bea
\Delta B_{\mu\nu\,M} &\sim& \xi^\rho {\cal H}_{\rho\mu\nu\,M}
\;,
\eea
upon applying the first order duality equation in now Euclidean external space.
As for invariance of (\ref{ExFTEuc}) under external diffeomorphisms, going through the 
original calculation \cite{Hohm:2013vpa},
the new sign in (\ref{DBEuc}) precisely compensates the additional sign arising from (\ref{MMMdEuc}) and 
the effect of a Euclidean external metric implying $\varepsilon_{\mu\nu\rho\sigma\tau}\varepsilon^{\mu\nu\rho\sigma\tau}=+120$.

\subsection{Parametrization of the generalized metric: M-theory, M$^*$-theory, and M'-theory}

With the Lagrangian (\ref{ExFTEuc}) uniquely fixed by internal and external generalized diffeomorphisms
together with the signatures of external and generalized internal space-time, the embedding of the higher-dimensional
theories is based on solving the section constraints and
establishing the dictionary between the ExFT fields and the components of the higher-dimensional
fields in analogy to (\ref{M27})--(\ref{compM2}).

The symmetric matrix ${\cal M}_{MN}$ in (\ref{ExFTEuc}) parametrizes the 
coset space ${\rm E}_{6(6)}/{\rm USp}(4,4)$. The identification of the fields from eleven-dimensional
supergravity is accomplished by solving the section constraints according to (\ref{break11D})
and choosing an explicit associated parametrization (\ref{M27})
with the last line now given by
\bea
{\cal M}_{\bar{m}n} &=& \frac1{24} \,|g|^{-2/3}\,  g_{mk} \,\epsilon^{klpqrs} \,c_{nlp}c_{qrs}-|g|^{-2/3}\,  g_{mn}\,\varphi
\;,
\nonumber\\
{\cal M}_{\bar{m}}{}^{kl} &=& -\frac1{6\,\sqrt{2}}\,|g|^{-2/3}\,g_{mn} \epsilon^{nklpqr}\,  c_{pqr}\;,
\qquad
{\cal M}_{\bar{m}\bar{n}} ~=~ -|g|^{-2/3}\, g_{mn}\;,
\label{compM1NC}
\eea
differing in the sign of the last component from (\ref{M27}).
The internal metric $g_{mn}$ is now of signature $(5,1)$ with negative determinant.
The remaining blocks of (\ref{M27}) are expressed in compact form through the matrix (\ref{tildeM27})
via
\bea
\tilde{\cal M}_{mn}  &=& |g|^{1/3}\,g_{mn} + \frac12\,|g|^{1/3}\,c_{mkp}c_{nlq} \,g^{kl} g^{pq}
\;,\nonumber\\
\tilde{\cal M}_{m}{}^{kl}  &=& -\frac1{\sqrt{2}}\,|g|^{1/3}\,c_{mpq} \,g^{pk} g^{ql}
\;,\qquad
\tilde{\cal M}^{kl,mn}  ~=~ |g|^{1/3}\,g^{m[k}g^{l]n}
\;.
\label{compM2NC}
\eea
Upon combining this parametrization with the proper dictionary for the ExFT $p$-forms ${A}_\mu{}^M$ and
${B}_{\mu\nu\,M}$, the Lagrangian (\ref{ExFTEuc}) reproduces full eleven-dimensional supergravity.
We refrain from spelling out all the details and give a simple illustration from the leading kinetic term of the 
vector fields ${A}_\mu{}^M$. Upon setting the scalars $c_{kmn}$ and $\varphi$ to zero, the Yang-Mills
term from (\ref{ExFTEuc}) reduces to
\bea
{\cal L}_{\rm vec} &=&
-\frac{1}{4}\,e\,{\cal M}_{MN}{\cal F}^{\mu\nu M}{\cal F}_{\mu\nu}{}^N
\nonumber\\
&=&
-\frac{1}{4}\,\sqrt{|g_{11D}|}
\left(
g_{mn}{\cal F}^{\mu\nu m}{\cal F}_{\mu\nu}{}^n
+g^{kl}g^{mn}{\cal F}^{\mu\nu}{}_{km}{\cal F}_{\mu\nu\,ln}
-|g|^{-1}\,g_{mn}\,{\cal F}^{\mu\nu \bar m}{\cal F}_{\mu\nu}{}^{\bar n}
\right)
\,,
\label{kineticVectorM}
\eea
where we have split vectors ${A}_\mu{}^M$ according to (\ref{break11D}).
The first term descends from the $11D$ Einstein-Hilbert term, the next two terms descend
from the kinetic term of the $11D$ four-form field strength.
The opposite sign of the last term is due to the sign appearing in the last component of (\ref{compM1NC})
and reflects the fact that it descends from dualization of two-forms $B_{\mu\nu\,\bar{m}}$ on Euclidean space.
With the internal metric $g_{mn}$ of signature $(5,1)$, it is straightforward to confirm that the 
signature of the matrix ${\cal M}_{MN}$ splits over the three terms
in (\ref{kineticVectorM}) as
\bea
(5,1)+(10,5)+(1,5) &=& (16,11)
\;,
\label{signMM1}
\eea
precisely as consistent with the coset ${\rm E}_{6(6)}/{\rm USp}(4,4)$.

A crucial difference with the standard ExFT reviewed in section~\ref{sec:review} above is due to the fact the
scalar matrix ${\cal M}_{MN}$ is no longer positive definite. As a result, the parametrization (\ref{compM1NC})
does not guarantee that the internal metric $g_{mn}$ extracted from the submatrix ${\cal M}_{\bar{m}\bar{n}}$
is non-degenerate. The dictionary with $D=11$ supergravity might thus break down even though the ExFT
configuration remains perfectly regular. 
This is comparable to the situation in \cite{Hohm:2011zr,Hohm:2011dv,Andriot:2011uh,Andriot:2012wx} in the O$(d,d)$ case.
Passing through such a degenerate point, one may reach a configuration in which the 
blocks (\ref{M27}) of the
matrix ${\cal M}_{MN}$
are parametrized as
\bea
{\cal M}_{\bar{m}n} &=& \frac1{24} \,|g|^{-2/3}\,   g_{mk} \,\epsilon^{klpqrs} \,c_{nlp}c_{qrs}-|g|^{-2/3}\,   g_{mn} \,\varphi
\;,
\nonumber\\
{\cal M}_{\bar{m}}{}^{kl} &=& -\frac1{6\,\sqrt{2}}\,|g|^{-2/3}\, g_{mn} \epsilon^{nklpqr}\,  c_{pqr}\;,
\qquad
{\cal M}_{\bar{m}\bar{n}} ~=~ |g|^{-2/3}\,  g_{mn}\;,
\label{compM1NC*}
\eea
for the last line, and
\bea
\tilde{\cal M}_{mn}  &=& |g|^{1/3}\,g_{mn} - \frac12\,|g|^{1/3}\,c_{mkp}c_{nlq} \,g^{kl} g^{pq}
\;,\nonumber\\
\tilde{\cal M}_{m}{}^{kl}  &=& \frac1{\sqrt{2}}\,|g|^{1/3}\, c_{mpq} \,g^{pk} g^{ql}
\;,\qquad
\tilde{\cal M}^{kl,mn}  ~=~- |g|^{1/3}\,g^{m[k}g^{l]n}
\;, 
\label{compM2NC*}
\eea
for the remaining blocks via the matrix (\ref{tildeM27}). In this parametrization, the internal metric $g_{mn}$ is of signature  $(4,2)$. 
As an illustration of its higher-dimensional origin, we again evaluate the Yang-Mills
term from (\ref{ExFTEuc}) at $c_{kmn}=0=\varphi$ to find
\bea
{\cal L}_{\rm vec} &=&
-\frac{1}{4}\,e\,{\cal M}_{MN}{\cal F}^{\mu\nu M}{\cal F}_{\mu\nu}{}^N
\nonumber\\
&=&
-\frac{1}{4}\,\sqrt{|g_{11D}|}
\left(
g_{mn}{\cal F}^{\mu\nu m}{\cal F}_{\mu\nu}{}^n
-g^{kl}g^{mn}{\cal F}^{\mu\nu}{}_{km}{\cal F}_{\mu\nu\,ln}
+|g|^{-1}\, g_{m n}\, {\cal F}^{\mu\nu \bar m}{\cal F}_{\mu\nu}{}^{\bar n}
\right)
\label{kineticVectorM*}
\eea
W.r.t.\ (\ref{kineticVectorM}), the last two terms have switched sign. With this parametrization of ${\cal M}_{MN}$,
keeping the same solution (\ref{break11D}) of the section constraint,
the theory (\ref{ExFTEuc}) thus describes a $D=11$ theory of signature $(5+4,2)=(9,2)$ in which the kinetic
term of the three-form, responsible for the last two terms in (\ref{kineticVectorM*}), has switched its sign.
This is precisely the field theory limit of Hull's M$^*$ theory \cite{Hull:1998ym}. It is defined such that
reduction on a timelike circle yields the IIA$^*$ supergravity, which we have thus also embedded into (\ref{ExFTEuc}).
Similar to (\ref{signMM1}), the signatures of the terms in (\ref{kineticVectorM*}) still add up to the same result
\bea
(4,2)+(8,7)+(4,2) &=& (16,11)
\;,
\label{signMM2}
\eea
fixed by the coset space ${\rm E}_{6(6)}/{\rm USp}(4,4)$.

Let us finally mention that the remaining eleven-dimensional supergravity proposed by Hull in \cite{Hull:1998ym},
the so-called $M'$ theory, is of signature $(6,5)$ and thus simply embeds into (\ref{ExFTEuc}) by
means of the dictionary (\ref{compM1NC})--(\ref{compM2NC}) with the flip $g_{mn}\rightarrow-g_{mn}$,
in order to account for an internal metric of signature $(1,5)$.

\subsection{Parametrization of the generalized metric: IIB and IIB$^*$}

Here, we briefly describe how IIB and IIB$^*$ supergravity are embedded in (\ref{ExFTEuc}) in a similar way.
Both are based on the IIB solution (\ref{breakIIB}) of the section constraint and amount to parametrizations 
of the scalar matrix ${\cal M}_{MN}$ that again differ from the E$_{6(6)}/{\rm USp}(8)$ parametrization worked
out in \cite{Baguet:2015xha} by a number of sign flips in order to account for a coset space of indefinite signature.
As an illustration, we simply state the Yang-Mills term from (\ref{ExFTEuc}), evaluated for vanishing forms
$b_{mn\,\alpha}$ and $c_{klmn}$.

In the IIB parametrization, this term reads
\bea
{\cal L}_{\rm vec} &=&
-\frac{1}{4}\,e\,{\cal M}_{MN}{\cal F}^{\mu\nu M}{\cal F}_{\mu\nu}{}^N
\nonumber\\
&=&
-\frac{1}{4}\,\sqrt{|g_{10D}|}
\Big(
g_{mn}
\,{\cal F}^{\mu\nu\,m}{\cal F}_{\mu\nu}{}^{n}
+
 g^{mn} m^{\alpha\beta}\,{\cal F}^{\mu\nu}{}_{m\alpha}{\cal F}_{\mu\nu\,n\beta}{}
\nonumber\\
&&{}
\qquad\qquad\qquad
-|g|^{-1}\,g_{mk}g_{ln}\,{\cal F}^{\mu\nu mn}{\cal F}_{\mu\nu}{}^{kl}
 -|g|^{-1}\,m^{\alpha\beta}\,{\cal F}^{\mu\nu}{}_\alpha{\cal F}_{\mu\nu\,\beta}
\Big)
\;,
\label{kineticVectorIIB}
\eea
with $g^{mn}$ and $m^{\alpha\beta}$ of signature $(4,1)$ and $(2,0)$, respectively.
The first two terms directly descend from the $D=10$ Einstein-Hilbert and three-form field
strength terms.
The last two terms are obtained from dualization of the $D=5$ two-forms that descend from 
the $D=10$ four-form and two-forms, respectively. Their minus signs account for the fact that 
this dualization has been performed w.r.t.\ the Euclidean external metric $g_{\mu\nu}$\,.
A quick count confirms that the 
signature of the matrix ${\cal M}_{MN}$ splits over the four terms in (\ref{kineticVectorM}) as
\bea
(4,1)+(8,2)+(4,6)+(0,2) &=& (16,11)
\;,
\eea
as determined by the coset ${\rm E}_{6(6)}/{\rm USp}(4,4)$.

In turn, the IIB$^*$ parametrization of ${\cal M}_{MN}$ gives rise to a Yang-Mills term that takes the form
\bea
{\cal L}_{\rm vec} &=&
-\frac{1}{4}\,e\,{\cal M}_{MN}{\cal F}^{\mu\nu M}{\cal F}_{\mu\nu}{}^N
\nonumber\\
&=&
-\frac{1}{4}\,\sqrt{|g_{10D}|}
\Big(
g_{mn}
\,{\cal F}^{\mu\nu\,m}{\cal F}_{\mu\nu}{}^{n}
+
 g^{mn} m^{\alpha\beta}\,{\cal F}^{\mu\nu}{}_{m\alpha}{\cal F}_{\mu\nu\,n\beta}{}
\nonumber\\
&&{}
\qquad\qquad\qquad
+|g|^{-1}\,g_{mk}g_{ln}\,{\cal F}^{\mu\nu mn}{\cal F}_{\mu\nu}{}^{kl}
 -|g|^{-1}\,m^{\alpha\beta}\,{\cal F}^{\mu\nu}{}_\alpha{\cal F}_{\mu\nu\,\beta}
\Big)
\;,
\eea
with $g^{mn}$ and $m^{\alpha\beta}$ of signature $(4,1)$ and $(1,1)$, respectively.
W.r.t.\ (\ref{kineticVectorIIB}), the sign of the third term has changed, precisely capturing
the sign flip of RR fields in the IIB$^*$ theory not accounted for by the signature change
of the axion-dilaton matrix $m^{\alpha\beta}$\,.

The ExFT (\ref{ExFTEuc}) with its different solutions to the section constraint and
different parametrizations of the scalar matrix ${\cal M}_{MN}$ thus captures the full IIA/IIB 
and IIA$^*$/IIB$^*$ supergravities together with the $D=11$ supergravities of exotic signature.
It would be very exciting to further explore the 
possibility of regular ExFT solutions that interpolate between different higher-dimensional theories,
along the lines of \cite{Dijkgraaf:2016lym,Blair:2016xnn,Berman:2019izh}.

\section{Generalized IIB supergravity}

We finish by sketching yet another application of exceptional field theory: the embedding
of the so-called generalized type II supergravities. Following the recent discovery of integrable 
deformations of the string sigma model \cite{Delduc:2013qra,Kawaguchi:2014qwa,Hollowood:2014qma},
it has subsequently been understood that (unlike for $\lambda$-deformations \cite{Borsato:2016zcf}) 
the action of (most of) the $\eta$-deformations do not give rise to solutions of the 
standard supergravity equations,\footnote{See, however, \cite{Hoare:2018ngg}.} but to a generalization thereof worked out in \cite{Arutyunov:2015mqj,Wulff:2016tju}. 
Solutions of these generalized type II equations are obtained by applying a generalized T-duality 
to a solution of the original supergravities however (in contrast to standard T-duality) with non-isometric linear dilaton. 
This suggests that the generalized supergravities should have a natural place in the 
manifestly duality covariant formulation of exceptional field theory, 
which is indeed the case, as we shall briefly review following~\cite{Baguet:2016prz}.
This provides yet a different scenario for combining solutions of the section constraint (\ref{break11D}), (\ref{breakIIB}),
with particular parametrizations of the ExFT fields.

Let us start from the IIB split of cordinates (\ref{breakIIB}) and in addition impose the
existence of a Killing vector field in the IIB theory, such that the coordinates further split as
$\{y^a\}=\{y^i, y^*\}$, $(i=1, \dots, 4)$, 
with $\partial_* \Phi=0$ for all fields of the theory. 
Next, the IIB solution of the section constraint is relaxed
by extending the internal coordinate dependence of fields to the 5 coordinates
\bea
\{y^i, \tilde{y}\}\,, \quad i=1, \dots, 4\;,
\label{yyy}
\eea
including a specific additional coordinate $\tilde{y}\equiv\tilde{y}_{*+}$ from among the $\{\tilde{y}_{a\alpha}\}$.
The section condition (\ref{sectionE6}) is still satisfied for the section defined by (\ref{yyy}).
As a final step, one imposes a particular
additional $\tilde{y}$-dependence on the ExFT fields,
according to a generalized Scherk-Schwarz ansatz of the form~\cite{Hohm:2014qga}
\bea\label{SSU}
{\cal M}_{MN} &=& U_{M}{}^{{K}}(\tilde{y})\,U_{N}{}^{{L}}(\tilde{y})\,{\cal M}_{{K}{L}}(x,y^i)\;, 
\nonumber\\
 g_{\mu\nu} &=& \rho^{-2}(\tilde{y})\,{ g}_{\mu\nu}(x,y^i)\;,\nonumber\\
  {A}_{\mu}{}^{M} &=& \rho^{-1}(\tilde{y}) (U^{-1})_{{N}}{}^{M}(\tilde{y})\,{ A}_{\mu}{}^{{N}}(x,y^i) \;, 
  \nonumber\\
  { B}_{\mu\nu\,M} &=& \,\rho^{-2}(\tilde{y})\, U_M{}^{{N}}(\tilde{y})\,{B}_{\mu\nu\,{N}}(x,y^i)
  \;,
 \eea
Here, the matrix $U_M{}^{{N}}$ lives in the $({\rm SL}(2) \times {\rm SL}(6))$ subgroup of E$_{6(6)}$
and is defined as a product of the two diagonal matrices
\bea
(U_{{\rm SL}(2)})_\alpha{}^{{\beta}} &=& 
\left(
\begin{array}{cc}
 U_+{}^+ & 0\\
0& U_-{}^- 
\end{array}
\right)~=~
\left(
\begin{array}{cc} \rho(\tilde{y}) & 0\\
0& \rho^{-1}(\tilde{y}) 
\end{array}
\right)\;,
\nonumber\\
(U_{{\rm SL}(6)})_{\hat a}{}^{{\hat b}}&=& 
\left(
\begin{array}{ccc
}U_i{}^j & 0&0\\
0& U_*{}^* & 0\\
0 & 0 & U_0{}^0 
\end{array}
\right)~=~
\left(
\begin{array}{ccc} \delta_i{}^j & 0&0\\
0& \rho(\tilde{y}) & 0\\
0 & 0 & \rho^{-1}(\tilde{y}) 
\end{array}
\right)\;,
\label{UU}
\eea
in the basis associated to the splitting (\ref{breakIIB}), (\ref{yyy}),
and with the scale factor $\rho$ given by a linear function $\rho(\tilde{y})=\tilde{y}+c$
of the extra coordinate.

Upon expanding the ExFT equations derived from (\ref{ExFTE6})
in the IIB parametrization of the fields, the coordinate dependence on $\tilde{y}$
entirely factors out, while the linear dependence of $\rho(\tilde{y})$ induces a
deformation of the IIB equations parametrized by the Killing vector field $K$~\cite{Baguet:2016prz}.
This precisely reproduces the results from \cite{Arutyunov:2015mqj,Wulff:2016tju}. 
It is important to note that
the choice of coordinates (\ref{yyy}) is equivalent (after rotation of the 27 coordinates) 
to selecting five coordinates among the $y^i$ in (\ref{break11D}).
Applying the same rotation to the IIA parametrization of ExFT fields such as (\ref{compM1}), (\ref{compM2})
would simply recover the IIA theory. This is a manifestation of the fact that the generalized IIB supergravity equations
can be obtained via T-duality from a sector of IIA supergravity. Since the framework of exceptional field theory
is manifestly duality covariant, the above sketched construction via (\ref{SSU}), (\ref{UU}) corresponds to
absorbing the effect of this duality into a rotation of the coordinates on the extended space.

\section{Conclusions}

We have reviewed the construction of exceptional field theory as the manifestly duality covariant formulation
of supergravity theories. Its dynamics is entirely determined by external and internal diffeomorphism invariance.
Upon solving the section constraints and choosing the proper parametrization of the ExFT fields, these equations
of motion reproduce the full higher-dimensional $D=11$ and IIB supergravity equations. What we have further
sketched in these proceedings is the potential of this formulation to provide a framework for embedding the magic
triangles of supergravity, the IIA$^*$/IIB$^*$ supergravities completing the orbit of supergravities under timelike dualities,
as well as the generalized type II supergravities emerging from integrable deformations.

Looking ahead, we may envision that ExFT provides solutions that transcend supergravity, despite the section constraints 
being everywhere satisfied. One may have 
a generalized metric or generalized frame that evolves through a phase that is singular from the 
viewpoint of ordinary geometry but perfectly regular from the viewpoint of generalized exceptional geometry.  
This approach towards `non-geometry', which is in the spirit of \cite{Andriot:2011uh,Andriot:2012wx} 
and for which new examples have been worked out in \cite{Berman:2019izh,Blair:2016xnn}, 
is quite different from the idea of relaxing the section constraints, which at least in the full string/M-theory must be viable too. 
It remains an exciting endeavor to explore the ultimate scope of exceptional field theory.

\subsection*{Acknowledgements}
We would like to thank the organizers of the Corfu Summer Institute 2018. 
The work of O.H.~is supported by the ERC Consolidator Grant ``Symmetries \& Cosmology".

%\bibliographystyle{JHEP}
%\bibliography{refs}

\providecommand{\href}[2]{#2}\begingroup\raggedright\endgroup

\end{document}